\documentstyle[epsf,prl,twocolumn,aps,floats]{revtex} 
\begin{document}
\topmargin = -2.0cm
\overfullrule 0pt

\twocolumn[\hsize\textwidth\columnwidth\hsize\csname
@twocolumnfalse\endcsname

\title{
Atmospheric neutrino observations and flavor changing interactions}

\author{M.\ C.\ Gonzalez-Garcia$^{1}$, 
M.\ M.\ Guzzo$^{2}$, 
P.\ I.\ Krastev$^{3}$, 
H.\ Nunokawa$^{2}$, \\
O.\ L.\ G.\ Peres$^{1}$, 
V. Pleitez$^{4}$, 
J.\ W.\ F.\ Valle$^{1}$ and 
R. Zukanovich Funchal$^{5}$ }
\address{\sl $^1$ Instituto de F\'{\i}sica Corpuscular -- C.S.I.C. \\
    Departamento de F\'{\i}sica Te\`orica, Universitat of Val\`encia \\
    46100 Burjassot, Val\`encia, Spain\\
    $^2$ Instituto de F\' {\i}sica Gleb Wataghin\\
    Universidade Estadual de Campinas, UNICAMP\\    
    13083-970 -- Campinas, Brazil \\
    $^3$ University of Wisconsin\\
    Madison, WI 53706, U.S.A. \\
    $^4$ Instituto de F\' {\i}sica Te\'orica \\
    Universidade Estadual Paulista\\
    R. Pamplona 145, 01405-900 S\~ao Paulo, Brazil \\
    $^5$ Instituto de F\'{\i}sica \\ 
       Universidade de S\~ao Paulo \\
    C.\ P.\ 66.318, 05389-970 S\~ao Paulo, Brazil. }
\maketitle
\vspace{.5cm}

\hfuzz=25pt

\begin{abstract} 
Flavor changing (FC) neutrino-matter interactions can account for the
zenith-angle dependent deficit of atmospheric neutrinos observed in
the SuperKamiokande experiment, without directly invoking neither
neutrino mass, nor mixing. We find that FC $\nu_\mu$-matter
interactions provide a good fit to the observed zenith angle
distributions, comparable in quality to the neutrino oscillation
hypothesis.  The required FC interactions arise naturally in many
attractive extensions of the Standard Model.
\end{abstract}
\pacs{PACS numbers: xxx}
\vskip2pc]

\newpage

Neutrinos produced as decay products in hadronic showers from cosmic
ray collisions with nuclei in the upper atmosphere ~\cite{flux} have
been observed by several detectors
\cite{Frejus,Nusex,Kamiokande,IMB,SuperKamiokande,Soudan}.  Although
the absolute fluxes of atmospheric neutrinos are largely uncertain,
the expected ratio $(\mu/e)$ of the muon neutrino flux ($\nu_\mu +
\bar{\nu}_\mu$) over the electron neutrino flux ($\nu_e+\bar{\nu}_e$)
is robust, since it largely cancels out the uncertainties associated
with the absolute flux.  In fact, this ratio has been calculated
~\cite{flux} with an uncertainty of less than 5\% over energies
varying from 0.1~GeV to 100~GeV. In this resides our confidence on the
long-standing atmospheric neutrino anomaly.

Although the first iron-calorimeter detectors in
Fr\'ejus~\cite{Frejus} and NUSEX~\cite{Nusex} reported a value of the
double ratio, R($\mu/ e$) = $(\mu/e)_{\rm data}/(\mu/e)_{\rm MC}$,
consistent with one, all the water Cherenkov detectors
Kamiokande~\cite{Kamiokande}, IMB~\cite{IMB} and
SuperKamiokande~\cite{SuperKamiokande} have measured R($\mu/ e$)
significantly smaller than one.  Moreover, not long ago, the Soudan-2
Collaboration, also using an iron-calorimeter, reported a small value
of R($\mu/ e$)~\cite{Soudan}, showing that the so-called atmospheric
neutrino anomaly was not a feature of water Cherenkov detectors.

Recent SuperKamiokande high statistics
observations~\cite{SuperKamiokande} indicate that the deficit in the
total ratio R($\mu/ e$) is due to the number of neutrinos arriving in
the detector at large zenith angles.  Although $e$-like events do not
present any compelling evidence of a zenith-angle dependence, the
$\mu$-like event rates are substantially suppressed at large zenith
angles.

The $\nu_\mu \to \nu_\tau$ \cite{SuperKamiokande,atm98} as well as the
$\nu_\mu \to \nu_s$ \cite{atm98,yasuda} oscillation hypothesis provide
an appealing explanation for this smaller-than-expected ratio, as they
are simple and well-motivated theoretically. This led the
SuperKamiokande Collaboration to conclude that their data provide good
evidence for neutrino oscillations and neutrino masses.

In this letter we give an alternative explanation of the atmospheric
neutrino data in terms of FC neutrino-matter interactions
\cite{wolfenstein,fy88,valle87,gmp91,kb97}. We show that even if 
neutrinos have vanishing masses and/or the vacuum mixing angle is
negligible, FC neutrino matter interactions can still explain the
SuperKamiokande data.

There are attractive theories beyond the SM where neutrinos are
naturally massless~\cite{gleb98} as a result of a protecting
symmetry, such as B-L in the case of supersymmetric $SU(5)$
models~\cite{FCSU5} and the model proposed in \cite{MV} or chiral
symmetry in theories with extended gauge structure such as $SU(3)_c
\otimes SU(3)_L \otimes U(1)_N$ (331) models \cite{331}. The simplest
example of this mechanism was first noted in an $SU(2)\times U(1)$
model proposed in Ref.~\cite{valle87,MV} where a singlet Dirac lepton
at the TeV scale is added sequentially to the SM in such a way that
neutrinos remain massless due to an imposed B-L symmetry.  The flavor
mixing amongst the massless neutrinos in the leptonic charged current
can not be rotated away despite neutrinos being massless.

In general, models of neutrino mass are a natural source of FC
neutrino-matter interactions.  Seesaw-type models of neutrino mass
have non-diagonal neutral current couplings of the $Z$ to mass
eigenstate neutrinos \cite{2227} that may lead to new FC
neutrino-matter interactions.  Models of radiative generation of
neutrino mass~\cite{zee.babu} typically contain additional  FC
neutrino-matter interactions from scalar exchanges. An example of this
class are the supersymmetric models with broken
$R$-parity~\cite{rpv}. Models with extended gauge structure, such as
$E(6)$ models \cite{gleb98}, may also lead to FC neutrino-matter
interactions.

Here we focus on a massless neutrino conversion scenario as an
explanation of the atmospheric neutrino data based on FC
neutrino-matter interactions which induce $\nu_\mu \to \nu_\tau$
transition.
Our results can be extended to the massive neutrino case. However, we
stress that the present atmospheric neutrino data does not necessarily
provide evidence for neutrino mass. Moreover, the existence of
attractive theories where FC neutrino-matter interactions do not imply
neutrino mass, makes this possibility especially elegant.
{}From a phenomenological point of view FC interactions of neutrinos can
induce flavor transitions when neutrinos travel through
matter~\cite{wolfenstein} irrespective of neutrino mass.  In both
massive~\cite{MS} and massless~\cite{valle87,gmp91} cases conversions
can be resonant, however the properties of the conversion are totally
different. Massless neutrino conversions would be energy-independent
and would affect atmospheric neutrinos as well as anti-neutrinos,
converting $\nu_\mu \to \nu_\tau$ together with $\bar\nu_\mu \to
\bar\nu_\tau$.  Remarkably the present double ratio data does not show
a significant energy dependence~\cite{SuperKamiokande,upgo}.

The presence of FC neutrino-matter interactions implies a non-trivial
structure of the neutrino evolution Hamiltonian in matter.  The
evolution equations describing the $\nu_\mu \to \nu_\tau$ transitions
in matter are given as~\cite{valle87,gmp91}:
\begin{eqnarray} &  i{\displaystyle{d}\over 
\displaystyle{dr}}\left( 
\begin{array}{c} \nu_\mu 
\\ \nu_\tau  \end{array} \right) = \hskip .3cm & \hskip-.1cm
\sqrt{2}\,G_F \left( \begin{array}{cc} 0 &  \epsilon_\nu n_f(r)
\\ \epsilon_\nu n_f(r)& \epsilon_{\nu} ' n_f(r) \end{array} \right)
\left( \begin{array}{c} \nu_\mu  \\ \nu_\tau 
\end{array} \right) ,
\label{motion} 
\end{eqnarray}
where, $\nu_a \equiv \nu_a (r)$, a=$\mu,\tau$ are the probability
amplitudes to find these neutrinos at a distance $r$ from their
creation position, $\sqrt{2}\,G_F n_f(r) \epsilon_\nu$ is the
$\nu_\mu+ f \to \nu_\tau + f$ forward scattering amplitude and
$\sqrt{2}\,G_F n_f(r) \epsilon_\nu '$ is the difference between the
$\nu_\tau - f$ and $\nu_\mu - f$ elastic forward scattering
amplitudes, with $n_f(r)$ being the number density of the fermions
which induce such processes.

The use of the FC $\nu_e$-matter interactions was previously suggested
in connection with the solar neutrino problem~\cite{gmp91,kb97}.
Recently an attempt was made~\cite{brooijmans} to extend this idea in
order to account also for the atmospheric neutrino data, but the fit
obtained in this paper is not as good as our atmospheric neutrino fit
or the solar fit in Ref. \cite{kb97}.  Moreover, some of the results
in Table III of Ref.~\cite{brooijmans} seem inconsistent.  On the
other hand Ref.~\cite{ma} includes exotic flavor-conserving $\nu_\tau$
interactions plus neutrino masses in order to account for the
atmospheric and LSND data, but without a detailed fit. We have decided
to postpone the detailed analysis (within the present scenario) of the
LSND, as well as of the solar neutrino data, for a future work.  Here
we show that FC $\nu_\mu$-matter interactions can explain the
atmospheric neutrino zenith angle anomaly, without introducing
neutrino masses and/or mixing.

For our phenomenological approach let us simply assume the existence
of a tree-level process $\nu_\alpha+ f \to \nu_\beta + f$ with
amplitude proportional to $\displaystyle g_{\alpha f}g_{\beta
f}/4m^2$, where $\alpha$ and $\beta$ are flavor indices, $f$ stands
for the interacting elementary fermion (charged lepton, $d$-like or
$u$-like quark) and $g_{\alpha f}$ is the coupling involved in the
vertex where a $\nu_\alpha$ interacts with $f$ through a scalar or
vector boson of mass $m$. The evolution equations which describe the
$\nu_\mu \to \nu_\tau$ transitions in matter may be written
generically as in Eq.~(\ref{motion}), where:
\begin{equation}
\epsilon^\prime_\nu = {|g_{\tau f}|^2 - |g_{\mu f}|^2 \over
4m^2\sqrt{2}\,G_F },\;\;\;\;\;\;
\mbox{and}\;\;\;\;\;\;
\epsilon_\nu   = {g_{\tau f} \cdot g_{\mu f} \over 4m^2\sqrt{2}\,G_F}. 
\label{epsilon}
\end{equation}
Note also that, in the absence of neutrino mass as we are assuming,
anti-neutrino transitions $\bar\nu_\mu\to \bar\nu_\tau$ are governed
by precisely the same evolution matrix in Eq.~(\ref{motion}).  We have
calculated the transition probabilities of $\nu_\mu~ \to \nu_\tau$
($\bar \nu_\mu \to \bar\nu_\tau$)
$P(\epsilon_\nu,\epsilon^\prime_\nu)$,
($P(\epsilon_{\bar\nu},\epsilon^\prime_{\bar\nu})$) as a function of
the zenith angle by numerically solving the evolution equation using
the density distribution in \cite{PREM} and a realistic chemical
composition with proton/neutron ratio 0.497 in the mantle and 0.468 in
the core \cite{BK}.  For the sake of simplicity we have assumed that
$\epsilon\equiv \epsilon_{\bar\nu}=\epsilon_\nu$ and $\epsilon^\prime
\equiv \epsilon^\prime_{\bar\nu}=\epsilon^\prime_\nu$, so there are
only two free parameters in the analysis.  We have used these
probabilities to compute the theoretically expected number of $\mu$-
and $e$-like events ($N_\mu$ and $N_e$) as a function of the two
parameters, $\epsilon$ and $\epsilon^\prime$, for each of the
5 zenith angle bins both for the sub-GeV and for the multi-GeV
SuperKamiokande data.  Following Refs. \cite{atm98,fogli2}, we fit
separately the $\mu$- and $e$-like events taking into account the
correlation of errors.  The calculated numbers of events, $N_\mu$ and
$N_e$, as functions of zenith angle, have been compared with the
535-days SuperKamiokande data sample in order to determine the allowed
regions of $\epsilon$ and $\epsilon^\prime$ from a $\chi^2$
fit. We set our normalization assuming that the relevant neutrino
interaction in the Earth is only with down-type quarks.  Any other
scenario can be obtained from our results by rescaling the
$\epsilon$-parameters.

\begin{figure}[ht]
\centering\leavevmode
\epsfxsize=\hsize
\epsfbox{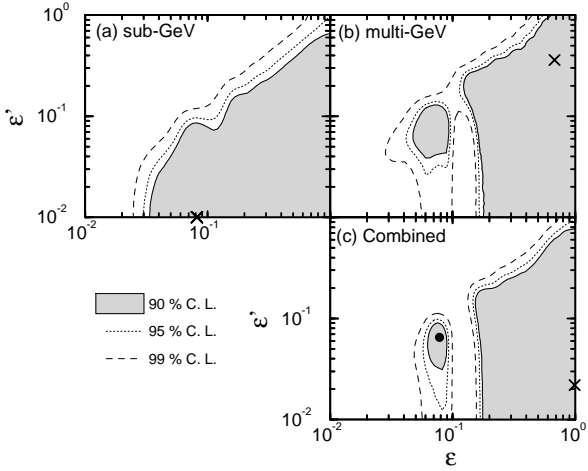}
\vglue -1cm
\caption{Allowed region for $\epsilon$ and $\epsilon'$ for
SuperKamiokande (a) sub-GeV (b) multi-GeV and (c) combined events in
the massless-neutrino scenario. The best fit points for each case is 
indicated by the crosses.}
\label{fig1}
\end{figure}

In Fig.(\ref{fig1}) we present a contour plot of the regions allowed
by the SuperKamiokande data.
The contour
plots (a), (b) and (c) correspond to the regions allowed by the
sub-GeV,  multi-GeV and combined SuperKamiokande data, respectively.
These contours are determined by the conditions $\chi^2 =
\chi^2_{min} + \Delta \chi^2$ where $ \Delta \chi^2 = 4.6, 6.0,
9.2$ for 90, 95 and 99 \% C.\ L., respectively.
\vglue 0.2cm
In the parameter region we have considered, i.e., $\epsilon$ and
$\epsilon'$ in the interval [0.01, 1.0], we found that $\chi^2_{min}$
= 6.3 and 6.4 for the sub-GeV and multi-GeV samples (8
d.~o.~f. corresponding to 10 data points minus two free
parameters). These minima are obtained for ($\epsilon, \epsilon') =$
(0.08, 0.01) and (0.68, 0.36), respectively, as indicated by the
crosses in Fig.(\ref{fig1}). 
For the combined case, $\chi^2_{min} = 14.7$ (18 d.o.f) for
($\epsilon,\epsilon') =$ (0.99, 0.02).
In the combined case the local best fit point ($\chi^2$ = 16.9 for
($\epsilon,\epsilon') =$ (0.08, 0.07) ) in the ``island'' determined
by the 90 \% C.\ L. curve is also indicated by a filled circle.  This
point is interesting because it still gives a good fit to the data
with a relatively small value for the FC parameter $\epsilon$.  We
find also that the ${\chi^2}$ is relatively flat along the
$\epsilon^\prime$ axis around the best fit point. The allowed regions
can be qualitatively understood in the approximation of constant
matter density. The conversion probability in this case is:

\begin{equation}
P(\nu_\mu \to \nu_\tau)= \frac{4\epsilon^2}{4\epsilon^2+\epsilon'^2}
\sin^2({1\over 2} \eta L), 
\label{Eq:prob}
\end{equation}
\noindent
where $\eta = \sqrt{4\epsilon^2+\epsilon'^2} \sqrt{2} G_F n_f$. 
For $n_f = n_d\approx 3n_e$ and $\epsilon' < \epsilon$, the oscillation 
length in matter is given by: 
\begin{equation}
L_{osc} = \frac{2\pi}{\eta} \approx 1.2\times 10^3 
\left[ \frac{2\ \mbox{mol/cc}}{n_e} \right]
\left[ \frac{1}{\epsilon} \right] \ \mbox{km}.
\label{osclength}
\end{equation}

{}From Eq.~(\ref{Eq:prob}) one can see that in order to have a large
transition probability one must be in the region $\epsilon^{\prime}
\lesssim \epsilon$ and $\eta\lesssim \pi/R_{\oplus}$. This last
condition leads to a lower bound on $\epsilon$.  The island in
Fig. 1.(b) corresponds to $\eta\sim\pi/R_{\oplus}$.

\begin{figure}[ht]
\centering\leavevmode
\epsfxsize=\hsize
\epsfbox{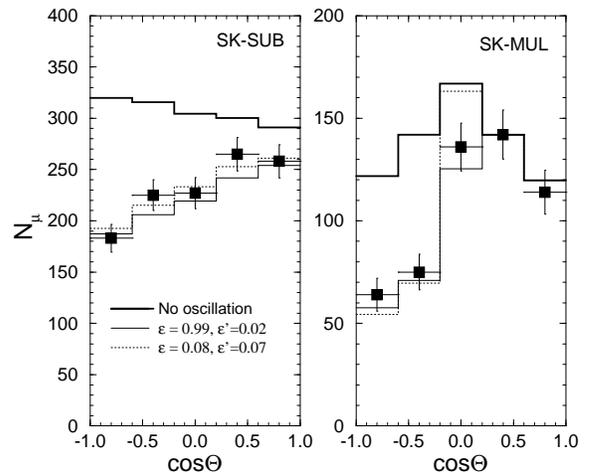}
\vglue -0.3cm
\caption{Best-fit zenith angle distributions in the massless-neutrino
FC scenario (thin-solid, dotted lines) versus no-oscillation 
hypothesis (thick-solid line). The SuperKamiokande data are 
indicated by the crosses.}
\label{fig2} 
\end{figure}

In Fig.(\ref{fig2}) we give the expected zenith angle distribution of
$\mu$-like sub-GeV events (a) and multi-GeV events (b) evaluated with
our Monte Carlo program for the best fit points determined above. Our
results clearly indicate an excellent fit for the $\mu$-like events
showing that they are highly depleted at $\cos\theta=-1$ with respect
to the SM prediction.  Note that, except for the assumption that the
FC $\nu_\mu$-matter interaction involves $d$-quarks, our result is
quite general, since we have not explicitly considered any particular
model as the origin of the FC neutrino-matter interaction.  Note that
$e$-like events are not affected by $\nu_\mu\to\nu_\tau$ transition.

What can we say about the required strength of the neutrino-matter
interaction in order to obtain a good fit of the observed data?
{}From our results and Eq.~(\ref{epsilon}) we see that for masses
$m\approx 200$ GeV we need at least $g_{\tau f} \cdot g_{\mu f} \sim
0.1$ for the the mixing term $\epsilon$.  Similarly our best fit
$\epsilon^\prime$ value implies $|g_{\tau f}|^2 - |g_{\mu f}|^2 \sim
0.1$.  While these values are relatively large, they are both
weak-strength couplings. Moreover they are consistent with present
experimental bounds, for example from universality of the weak
interaction which is manifestly violated by Eq.~(\ref{motion}).

For the purpose of illustrating this explicitly let us consider for
the moment the supersymmetric model with broken $R$-parity \cite{rpv}
as a way to parameterize the FC neutrino-matter interaction. In this
case the FC $\nu_\mu$-matter interactions are mediated by a scalar down-type
quarks, $\tilde{d_j}$, so that we need only to check the couplings
where a $d$-quark and a $\mu$- or $\tau$-neutrino is involved,{\it
i.e} $g_{id} \approx \lambda^\prime_{ij1}$, $i=2,3$. $\lambda_{ijk}'$
are the coupling constants in the broken $R$-parity superpotential
$\lambda_{ijk}' L_iQ_jD^c_k$, where $L$, $Q$ and $D$ are standard
superfields, and  $4 \sqrt{2} G_F \epsilon = |{\displaystyle
\sum_j}\lambda_{3j1} '\lambda_{2j1} '/\tilde{m}_{\tilde d j}^2|$.

Constraints on the magnitude of such FC interactions in broken
$R$-parity models have been given in \cite{giudice}. The precision
tests imply that individually these couplings are not severely
constrained.  The most stringent limit to the values of the relevant
FC quantities comes from limits on the FC tau decay BR($\tau^-\to
\rho^0+\mu^-) < 6.3 \times 10^{-6}$~\cite{pdg} which implies that
$|{\displaystyle \sum_j}\lambda_{3j1}' \lambda_{2j1} ' (100$
GeV$/\tilde{m}_{\tilde u j})^2 |< 3.1 \times 10^{-3}$.  Although a
certain degree of fine-tuning is needed in order to verify this
constraint we find that the required strength of FC $\nu_\mu$-matter
interaction is consistent with all present data. One should note that
broken $R$-parity supersymmetric models typically lead to neutrino
masses which could be large. However one may suppress them via
fine-tuning or some additional symmetry.

In summary, we have demonstrated that flavor changing
$\nu_{\mu}$-matter interactions can account for the zenith-dependent
deficit of atmospheric neutrinos observed in the SuperKamiokande
experiment, without directly invoking neutrino masses and mixing. It
provides a fit of the observations which is significantly better than
the no-oscillation hypothesis and of similar quality as the usual
$\nu_\mu \to \nu_\tau$ oscillation hypothesis.  The required FC
interaction can arise in many attractive extensions of the SM and is
consistent with all present constraints.

The above FC mechanism can also be tested at future Long Baseline
experiments. From Eq.~(\ref{Eq:prob}), using $n_e \sim 2$ mol/cc, $\epsilon
\sim 1 $ (0.1) and $\epsilon' < \epsilon $, for the planned K2K
experiment ~\cite{K2K} one gets $P(\nu_\mu \to \nu_\tau) \sim 0.4$
(0.004) while for MINOS~\cite{MINOS} one finds $P(\nu_\mu \to
\nu_\tau) \sim 0.9$ (0.2).

The existence of a massless neutrino explanation of the atmospheric
neutrino anomaly may play an important theoretical role in
model-building, especially if one wants to account for all other hints
for non-standard neutrino properties, namely the solar neutrino data,
the LSND result, and the possible role of neutrinos as dark matter in
the Universe \cite{gleb98}.



We are grateful to M. Drees, Y. Grossman and Y. Nir for useful
comments.  This work was supported by Spanish DGICYT under grant
PB95-1077, by the European Union TMR network ERBFMRXCT960090, by NSF
grant \#PHY96-05140, and by Brazilian funding agencies CNPq, FAPESP
and the PRONEX program.  \vglue -0.4cm

\vglue -0.2cm

\end{document}